\documentclass[onecolumn,showpacs,preprintnumbers,amsmath,amssymb]{revtex4}

\usepackage{graphicx}

\newcommand{\xa}{$X_A$}
\newcommand{\xb}{$X_B$}

\newcommand{\na}{$N_a$}
\newcommand{\nb}{$N_b$}

\newcommand{\da}{$D_a$}
\newcommand{\db}{$D_b$}

\begin{document}

\title{
Direct observation of acoustic phonon mediated relaxation 
between coupled exciton states in a single quantum dot molecule
}
\author{T. Nakaoka$^{1,2}$,
H. J. Krenner$^1$, E. C. Clark$^1$, M.~Sabathil$^1$, M.~Bichler$^1$, Y.~Arakawa$^2$, G.~Abstreiter$^1$, and J.~J.~Finley$^1$}
\affiliation{
$^1$ Walter Schottky Institut and Physik Department, Technische Universit\"at M\"unchen, Am Coulombwall 3, 85748 Garching, Germany\\
$^2$ Research Center for Advanced Science and Technology, University of Tokyo, 4-6-1 Komaba, Meguro-ku, Tokyo 153-8505, Japan \\
}

\date{\today}


\begin{abstract}
We probe acoustic phonon mediated relaxation between tunnel coupled exciton states in an individual quantum dot molecule in which the inter-dot quantum coupling and energy separation between exciton states is continuously tuned using static electric field. Time resolved and temperature dependent optical spectroscopy are used to probe inter-level relaxation around the point of maximum coupling. The radiative lifetimes of the coupled excitonic states can be tuned from $\sim2$ ns to $\sim10$ ns as the spatially \textit{direct} and \textit{indirect} character of the wavefunction is varied by detuning from resonance. Acoustic phonon mediated inter-level relaxation is shown to proceed over timescales comparable to the \emph{direct} exciton radiative lifetime, indicative of a relaxation bottleneck for level spacings in the range $\Delta E\sim3-6$ meV.
\end{abstract} 


\maketitle
The charge and spin degrees of freedom of localized charges in quantum dot (QD) nanostructures currently attract significant interest due to their potential to form quantum bits for an inherently scalable quantum processor based on solid-state hardware.  The implementation of such coherent devices using QDs can be traced to the strong three dimensional confinement that gives rise to relatively long coherence times for charge\cite{zrenner,hayashi} and spin\cite{kroutvar,lossnano} excitations when compared to the time required for quantum state manipulation.  Very recently, advances in QD fabrication have allowed the realization of self-assembled double dot nanostructures in which the quantum coupling between electronic states, and the spatial localization of the exciton wavefunctions, can be continuously varied by varying the potential applied to a metal gate contact.\cite{krenner,krenner2,ortner2}  Such tunable QD-molecules (QDMs) may allow ultrafast coherent manipulation using electro-optical techniques and may exhibit \textit{conditional} coherent dynamics.\cite{VilasBoas}
\\
A main obstacle to the realization of coherent QDM devices is decoherence caused by coupling to phonons.\cite{climente,muljarov,borri01,borriQDM} In pioneering work by Bockelmann\cite{bockelmann1} phonon mediated relaxation between discrete electronic states in QDs was predicted to occur over timescales comparable to the radiative lifetime, an effect commonly termed the phonon \emph{bottleneck}. However, experiments quickly revealed little or no evidence for bottleneck effects for level spacings of a few $10$´s of meV.  These findings were explained by the participation of multi phonon processes \cite{sakaki}, Coulomb scattering \cite{ohnesorge} or polaron formation.\cite{polaron2} In contrast, for level spacings of only a \textit{few} meV, acoustic phonons have been shown to be the dominant source of decoherence for both single \cite{fujisawa02,borri01} and vertically coupled\cite{ortner1,hayashi,borriQDM,climente,muljarov} quantum dots.  In this case the inter-level scattering rate can be strongly suppressed \textit{or} enhanced, depending on the spatial extent of the carrier wavefunction.\cite{climente} Therefore, experimental information on the phonon mediated coupling of the discrete states of a QDM is of fundamental importance for comparison with theory.
\\
In this paper we probe acoustic phonon mediated inter-level relaxation between molecular-like coupled single exciton ($e+h$) states in an individual self-assembled QDM.  The inter-dot tunnel coupling is controlled by varying the axial electric field, enabling us to tune the coupled state energy splitting in the range $\Delta E\sim 3$ - $6$ meV and control the mixing between spatially \emph{direct} ($e$ + $h$ localized in upper dot) and \textit{indirect} ($e$ in lower dot, $h$ in upper dot) excitons. By combining continuous wave (CW), time resolved and temperature dependent spectroscopy we demonstrate the ability to control the exciton spontaneous emission lifetime and probe acoustic phonon mediated relaxation as the level spacing and spatial character of the exciton wavefunction is varied. 


\begin{figure}
\includegraphics[width=\textwidth]{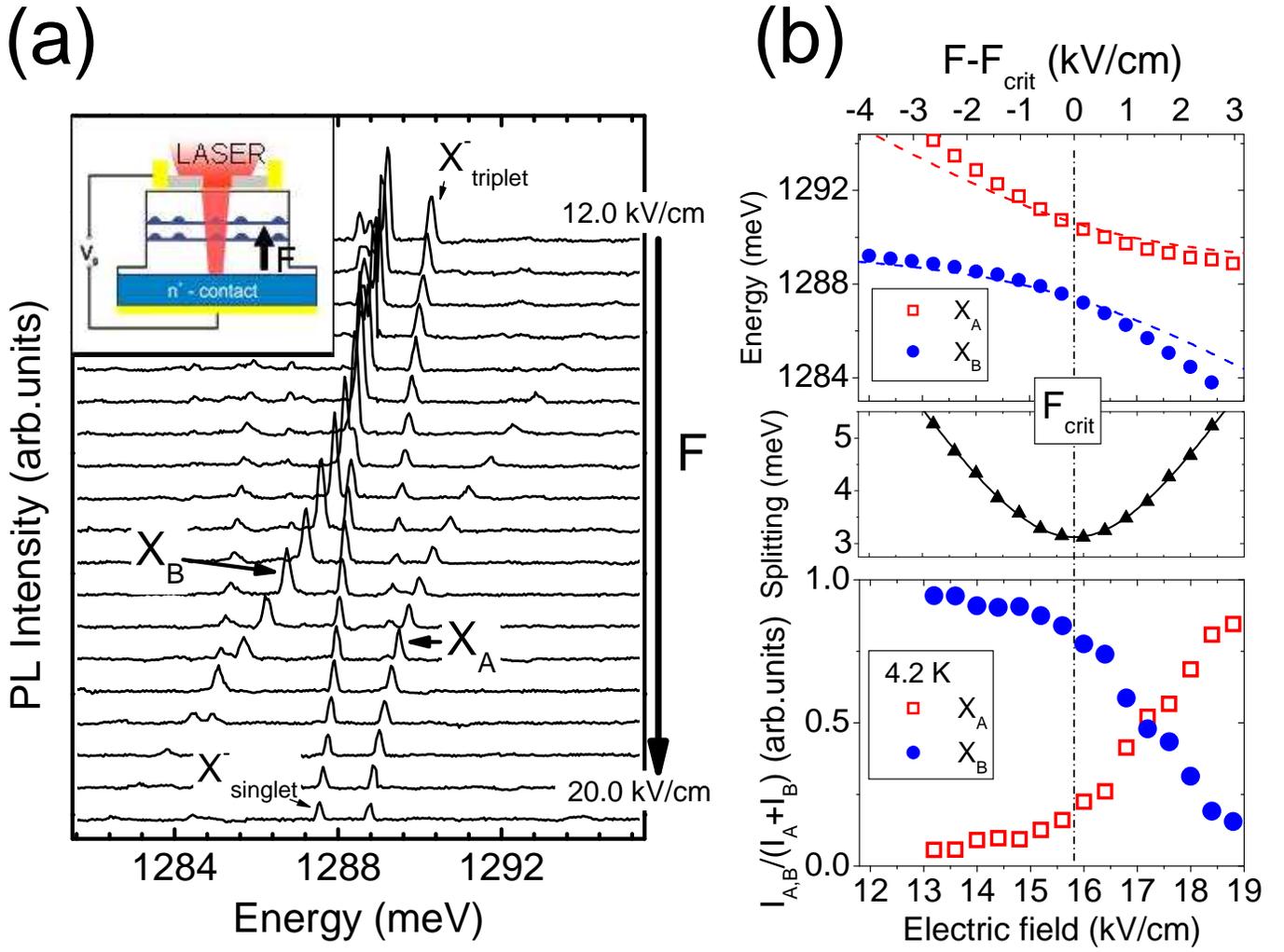}
\caption{(color online) (a) PL spectra of a single QDM as a function of the electric field.  
(Inset) schematic of the $n-i$ Schottky diodes studied.(b) Upper panel: Experimental (symbols) and calculated (dashed lines) exciton energies
for \xa \ and \xb.  
Middle panel: Energy splitting between \xa \ and \xb \ 
and fitting (see text).
Lower panel: Relative intensities $I_{A,B}/(I_A + I_B)$ and the resonant field (dashed line).
}
\label{fig1}
\end{figure}

%
The QDMs studied consisted of two layers of stacked InGaAs QDs with a nominal inter-dot separation of $d=10$ nm. The molecules were embedded in an $n-i$ Schottky photodiode structure to control the static electric field $(F)$ along the growth axis and Al shadow masks were defined to optically isolate single QDMs. A schematic of the device structure is presented in the inset of Fig. 1(a). Further details of the InGaAs-GaAs QDM epitaxial layer sequence and sample fabrication can be found elsewhere.\cite{krenner2,krenner}  
Time resolved studies were performed using a gated Si-APD and time-correlated single photon counting techniques that provided a temporal resolution of $\lesssim0.2$ ns. Carriers were photogenerated by a mode-locked Ti:Sapphire laser tuned at 1.48 eV to excite the QDM wetting layer that delivers $\sim$2 ps duration pulses with an energy of 6 fJ. 
\\
A series of time integrated photoluminescence (PL) spectra recorded as a function of electric field ($F$) are presented in the main panel of Fig. 1(a). A clear anti-crossing between two transitions, labeled $X_A$ and $X_B$, can clearly be observed.  These two peaks, which will be the focus of this paper, arise due to the coupling of spatially direct and indirect neutral excitons as discussed in our previous publications.\cite{krenner,krenner2}  Other peaks observed, even at the lowest excitation powers investigated ($P_{ex}$ $\sim$ 0.1 W/cm$^2$) also arise from the same QDM and arise from \textit{negatively} charged excitons as discussed in ref. \cite{krenner3}.  At a low field ($F \lesssim$ 14 kV/cm) $X_B$ is predominantly an unmixed state with spatially direct character.  Thus, it shifts weakly with field and has a high electron-hole overlap and, consequently, a large oscillator strength.  In contrast, \xa \ is a state with predominantly indirect character over this low field range and, thus, shifts strongly with $F$ and has a weak oscillator strength.\cite{krenner} As the electric field increases, \xa \ and \xb \ shift closer to one another and become increasingly mixed until, for $F_{crit}\sim 15.8$ kV/cm, they anticross.  At this point the \emph{electron} component of the exciton wavefunction hybridizes over both dots to produce exciton states with bonding (\xb~) and anti-bonding (\xa~) character.  
The measured peak positions are compared with effective mass calculations of the coupled exciton states that include strain, quantum tunneling coupling, and Coulomb interaction for a model QDM with $d = 11$ nm. \cite{krenner2} The results of this calculation are shown by the dashed lines on Fig 1(b) (top panel), reproducing well both the field dependence of the transition energies of \xa~and \xb~and the coupling energy at $F_{crit}$ marked by a vertical line. Below, we use these effective mass calculations to obtain the $F$-dependence of the relative oscillator strengths of \xa~and \xb~and compare them to the results of our time resolved and CW spectroscopy. As expected for a coupled quantum system, the energy splitting between \xa~and \xb~ is well represented by $\Delta E= \sqrt{ (2\Delta E_{e-e})^2 + (\delta_{e-e}\cdot (F - F_{crit}))^2}$ with $\delta_{e-e}\cdot(F - F_{crit})$ being the detuning from the point of maximum coupling [Fig. 1(b) - middle panel].
\\
The lower panel of Fig. 1(b) shows the relative intensities $I_{A,B}/(I_A+I_B)$ of the upper and lower exciton branches as the resonance field is traversed. Clearly, the field at which the two exciton branches have equal relative intensities is \textit{larger} than the critical electric field of $F_{crit}$=15.8 kV/cm. This observation contrasts strongly with the simple expectation that the radiative rates and populations are similar for both states at $F_{crit}$.\cite{similar-recom} Moreover, it indicates the presence of a relaxation process from \xa~to \xb~ which is active over timescales comparable to the radiative lifetime. We note that the observed intensity ratio of $I_{A}/I_{B}\sim 1:5$ does not reflect a thermal population distribution since only the lower exciton branch would be significantly populated 
$[\exp(-2\Delta E_{e-e}/k_{B}T)< 10^{-4}]$.  

\begin{figure}
\includegraphics[width=\textwidth]{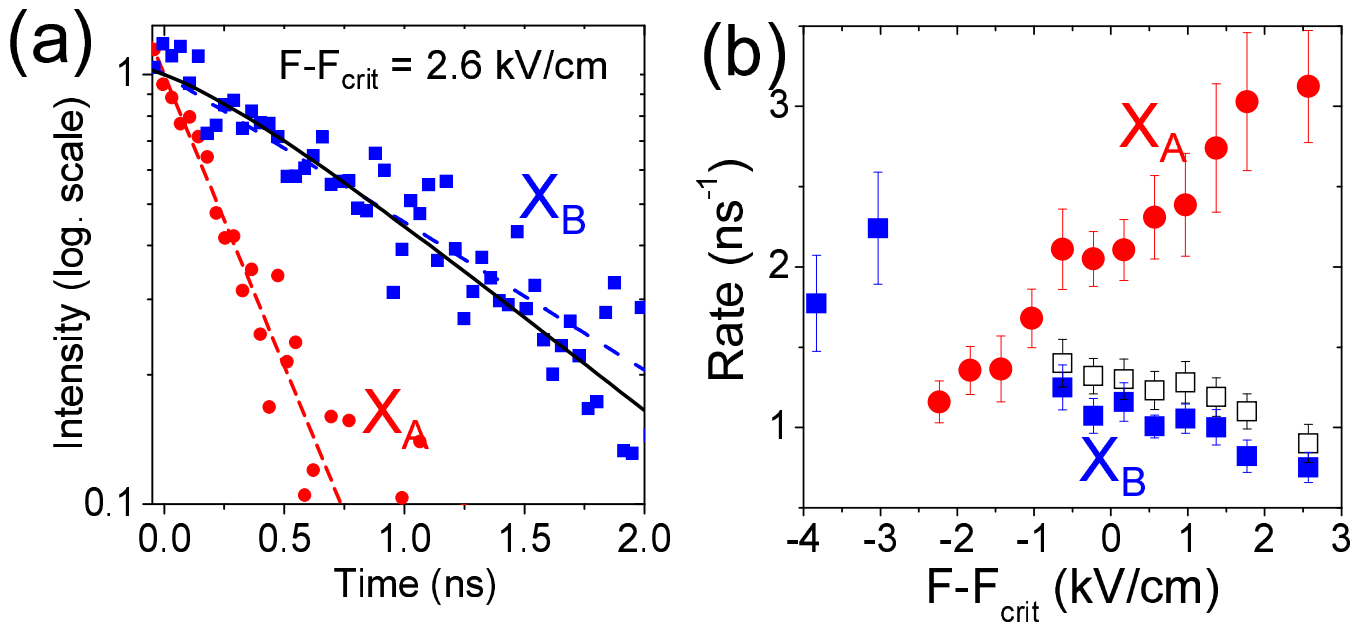}
\caption{
(color online) (a) Normalized PL decay curves for \xa \ and \xb,
measured at $F-F_{crit}=2.6 $kV/cm.
Dashed and the solid lines are fits
 with a mono-exponential function and 
a bi-exponential function, respectively.
(b) The PL decay rates for \xa \ ($D_a$)
(solid circles) and \xb \ (solid squares) obtained by the mono-exponential fittings.
Total population decay rates for \xb \ ($D_b$) 
(open squares) 
based on the rate equation model discussed in the text
are also shown.}
\label{fig2}
\end{figure}

In order to obtain further information about the nature of the relaxation process we performed time resolved spectroscopy on the two exciton branches as a function of electric field. Typical PL decay traces recorded for \xa \ and \xb \ are presented in Fig. 2(a) for an electric field \emph{detuned} ($F-F_{crit}=+2.6$ kV/cm) from the critical field.  Under these conditions \xb~exhibits a much slower decay than \xa~since the spatial overlap of the electron and the hole is much weaker for the mixed exciton state due to its stronger fraction of indirect character. 

Before entering into detailed analysis we present a summary of the decay lifetimes as a function of positive and negative detuning, either side of $F_{crit}$.   From fitting the observed decay transients using single exponential decay functions we obtain the decay rates for \xa \ and \xb \ presented in Fig. 2(b).  With increasing the electric field the decay rate measured for \xa \ increases by a factor of $\sim3\times$ whilst the \xb \ decay rate decreases by a similar factor.  This distinct anti-correlation between the field dependent PL-decay rates recorded for \xa \ and \xb \ demonstrates clearly that the electric field \emph{tunes} the direct-indirect character of the two mixed excitonic states.  States with strong \emph{indirect} character have lower decay rates, increasing as the \emph{direct} admixture becomes larger.  

\begin{figure}
\includegraphics[width=\textwidth]{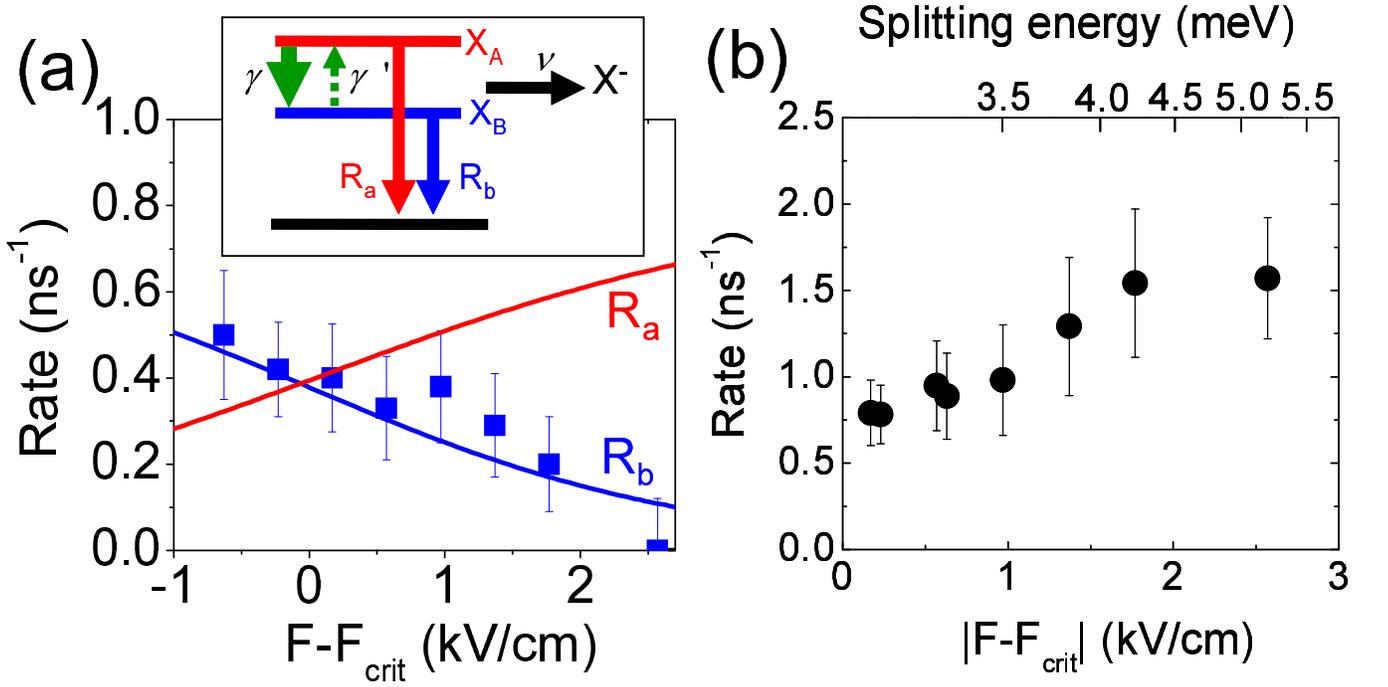}
\caption{
(color online) (a)
Experimental (solid squares) and calculated (solid line) 
radiative recombination rates
plotted as a function of $F$. (Inset) schematic processes for radiative recombinations of the exciton states \xa \ and \xb, relaxation between them, and electron capture. 
(b) Phonon relaxation rate $\gamma=\gamma_0$ at 4.2 K from \xa \ to \xb \
obtained from the time-resolved PL and rate-equation model.}
\label{fig3}
\end{figure}

Surprisingly we find that \xa \ and \xb \ do \emph{not} have the same decay rate at $F_{crit}$ despite our simple expectation that the two states should carry similar oscillator strength.  Instead, the observation of a different decay rate for \xa \ and \xb \ at $F_{crit}$ strongly indicates the existence of a non-radiative relaxation channel from the antibonding (\xa) to the bonding state (\xb) with a rate comparable to the pure radiative lifetime.  In order to quantify the relaxation rate and radiative lifetime, we consider a simplified rate equation model illustrated by the 3 level system presented in the inset of Fig. 3(a). In this model, the two neutral exciton levels \xa \ and \xb \ decay radiatively with electric field dependent rates $R_a$ and $R_b$, respectively. The non-radiative relaxation between the upper and lower exciton branches are taken into account by down ($\gamma$) and up ($\gamma'$) scattering rates due to phonon emission.  In addition, a phenomenological electron capture process with rate $\nu$ is included due to the prominence of negatively charged excitons in the PL spectra presented in Fig. 1(a).  The rate equations for the population of \xa \ (\na) and \xb \ (\nb) subject to the processes discussed above can be set in the form
 
\begin{eqnarray} 
\label{eq:rate}
\frac{dN_a}{dt} & = & -R_aN_a - \gamma N_a + \gamma 'N_b - \nu N_a + G, \\
\frac{dN_b}{dt} & = & -R_bN_b - \gamma 'N_b + \gamma N_a - \nu N_b + G,
\end{eqnarray}
where $G$ is the neutral exciton generation rate which is the same for \xa \ and \xb. 
In order to simplify further our analysis, we note that $\gamma' \ll  \gamma=\gamma_0$ at 4.2 K since the phonon population is negligible ($< 10^{-4}$) and solve the rate equations to obtain the time dependence of \na \ and \nb, viz.
\begin{eqnarray}
N_a & = & \exp(-D_at) \\
N_b & = & (1+\alpha)\exp(-D_bt)-\alpha \exp(-D_at).
\end{eqnarray}
Here $D_a = R_a + \nu + \gamma_0$ and $D_b = R_b + \nu $ are the measured decay rates for \xa \ and \xb \ and $\alpha = \gamma_0/(D_a - D_b)$.  The decay rates extracted from a least squares fitting of mono- ($N_a$) and bi-exponential ($N_b$) transients are presented in Fig. 2(b) from which we estimate $\gamma_0\sim 0.6 \pm 0.5$ ns$^{-1}$.  Here, the error is large as $\gamma_0$ is extracted from the relative amplitudes of the bi-exponential fit. Thus, we also estimated $\gamma_0$ by measuring the difference between \da \ and \db \ [Fig. 2(b)] at $F_{crit}$.  This provides a direct measure of the inter-level relaxation rate since $R_a(F_{crit})\approxeq R_b(F_{crit})$ providing that the charged exciton formation rate is the same for both \xa \ and \xb \ .  This approach yields $\gamma_0(F_{crit})=D_a(F_{crit}) - D_b(F_{crit}) \sim 0.7\pm 0.3$ ns$^{-1}$, in good agreement with the value obtained by fitting the bi-exponential decay transient.%
\\
In order to investigate the $F$-dependence of $\gamma_0$ we need to estimate $\nu$, the electron capture rate. To do this we focus on the field range $-1$ kV/cm $<(F-F_{crit})< 3$ kV/cm, since the charged excitons are pronounced at lower field and \db \ can be directly measured in this region.  We expect that the recombination rate of the \textit{indirect} exciton $R_b$ is much smaller than $\nu$ in the high field range ($F > 18$ kV/cm), since \xb \ is much weaker than other lines arising from charged exciton recombination [Fig. 1(a)]. We then estimate $\nu = 0.9$ ns$^{-1}$ from \db, constant in this narrow field range since the field dependence of the charged exciton intensity is much weaker than the neutral exciton transitions. Using this value for $\nu$, we deduce the radiative recombination rate $R_b$.  The results of this analysis is presented in Fig. 3(a) and compared with the calculated recombination rate obtained in the dipole approximation, i.e. $R={n_{r}e^2E_{exc}E_P}\left| \langle f_e|f_h\rangle\right| ^2/{2\pi \epsilon m_0c^3\hbar^2}$ where $n_r=3.4$ is the refractive index, $E_p=25.7$ eV is the Kane matrix element, $E_{exc}$ is the emission energy, and $\langle f_e|f_h\rangle$ is the overlap integral of the envelope functions obtained from the effective mass calculations used to fit the transition energies in Fig. 1(b).  Fairly good quantitative agreement is obtained between these calculations and the values of $R_b$ extracted from our rate equation analysis, strongly supporting the general validity of the analysis.  The radiative lifetime can be tuned from $\sim 2$ to $\sim 10$ ns as the fraction of direct character in the mixed exciton state is tuned, in good quantitative agreement with previous reports for vertically stacked QDs.\cite{bardot,gerardot}  From the extracted values of $R_b$ and $\nu$ we obtain the electric field dependence of $\gamma_0$ presented in Fig. 3(b). Typical values are of the order of $\gamma_0\sim1$ ns$^{-1}$ for a level separation $\Delta E= 3-4$ meV, consistent with recent theory for phonon mediated relaxation in QDMs with properties similar to those investigated here.\cite{muljarov,climente}

\begin{figure}
\includegraphics[width=\textwidth]{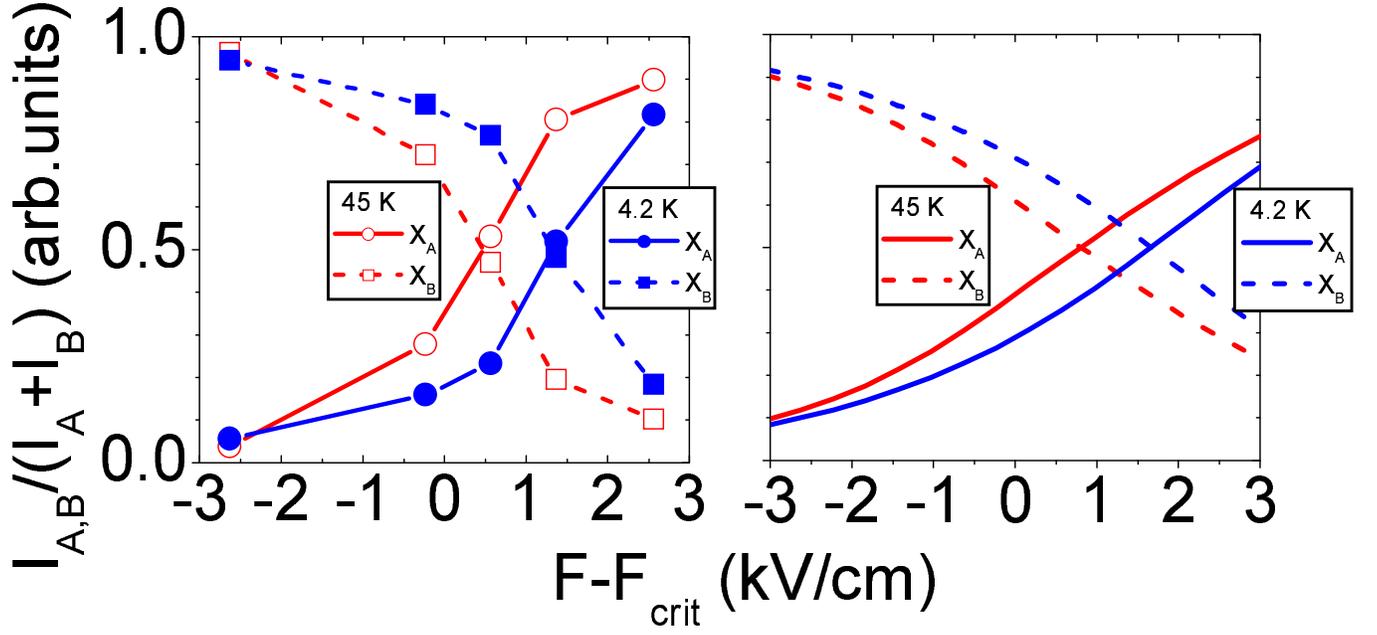}
\caption{(a) Experimental and (b) calculated 
relative intensities of \xa \ and \xb \ at 4.2 K and 45 K.
}
\label{fig4}
\end{figure}

At elevated temperatures both relaxation and activation processes determine the steady state populations $N_a$ and $N_b$ and, thus, the resulting intensities of \xa \ and \xb \ ($I_{A}$ and $I_{B}$). Fig. 4(a) shows the measured field dependence of the relative intensities at 4.2 K 
($\Delta E\gg k_BT$, $\gamma \gg \gamma^{'}$) and 45 K 
($\Delta E \sim k_BT$, $\gamma\sim \gamma^{'}$). 
At 45 K, $I_{A}\sim I_{B}$ for $F\sim F_{crit}$ as expected since $R_a\sim R_b$ at $F_{crit}$ [Fig 3(b)] and the activation process results in $N_a\sim N_b$. This processes is reproduced by our rate equation model with single phonon scattering processes. Solving eqns. (1) and (2) in steady state we obtain the PL intensity ratio for a CW excitation at zero temperature, i.e.
 
\begin{eqnarray}
\frac{I_{A}}{I_B}= \frac{1+(\nu + 2\gamma')/R_b}{1+(\nu + 2\gamma)/R_a}. 
\end{eqnarray}

At finite temperature, one-phonon emission (absorption) rates are given by 
$\gamma(T)=\gamma_0\{ N_{ph}(T)+1\}$ $[\gamma'(T)=\gamma_0 N_{ph}(T)]$, where $N_{ph}(T)$ is the phonon occupation factor.  The intensity ratios and their temperature dependence obtained by substituting the calculated $R_a$, $R_b$ and the estimated $\nu $ into the above equation are presented in Fig. 4(b), producing fairly good qualitative agreement with experiment.  In particular, the shift of the equal intensity point towards $F\sim F_{crit}$ is reproduced.  The broad agreement between the intensity ratios obtained from the CW PL measurement and calculation strongly supports the validity of our rate equation analysis and the assumptions made. %
In summary, we have demonstrated the existence of inter-level phonon relaxation between the molecular-like exciton states in an individual self-assembled QDM. The relaxation rate is found to be comparable to the radiative recombination rates of the direct excitons demonstrating the existence of a phonon bottleneck in the meV range.Furthermore, we tune the radiative recombination rates of a single molecule as the fraction of the direct exciton character is varied.  The observation and the calculation of the thermal evolutions of the exciton populations suggests a dominance of one-phonon processes for the inter-level relaxation in the QDM, with an intrinsic rate $\sim0.5-1.5$ ns$^{-1}$.
%


This work was supported in part by the IT program (RR2002) 
from the MEXT (Japan) and by the DPG via SFB631 (Germany).




\begin{references}
\bibitem{similar-recom}
Our calculation gives the recombination-rate difference of 4 \% 
between $R_a$ and $R_b$ at $F_{crit}$.
\end{references}
\end{document}